%% file: Phase_Space_SuperResolution_V2_ArXiV.tex
\documentclass[10pt]{IEEEtran}
\onecolumn
\usepackage[ruled,vlined]{algorithm2e}
\usepackage{url}

\include{ABPreICASSP2k15}

\include{BoxedEquations}

\begin{document}

\par\noindent {\LARGE\bf Super--Resolution in Phase Space \par}

{\vspace{4mm}\par\noindent
Ayush Bhandari$^\dag$, Yonina Eldar~$^\ddag$ and Ramesh Raskar$^\dag$
\par\vspace{2mm}\par}

{\vspace{2mm}\par\it
\noindent $^\dag$ Media Laboratory, Massachusetts Institute of Technology\\
$\phantom{^{\dag}}$ Cambridge, MA 02139--4307 USA. \par}

{\vspace{2mm}\par\it
\noindent $^\ddag$ Dept. of Electrical Engineering\\
 $\phantom{^{\dag}}$ Technion--Israel Institute of Technology, Haifa 32000, Israel.\par}

{\vspace{2mm}\par\noindent $\phantom{^{\dag,\ddag}}$\rm 
Email: \textrm{ayush@MIT.edu $\bullet$ \ yonina@ee.technion.ac.il $\bullet$ \ raskar@MIT.edu} \par}

{\vspace{6mm}\par\noindent\hspace*{5mm}\parbox{150mm}{\small \textbf{Abstract: }
This work \com{considers} the problem of super--resolution. The goal is to resolve a Dirac distribution from knowledge of its discrete, low--pass, Fourier measurements. Classically, such problems have been dealt with parameter estimation methods. Recently, it has been shown that convex--optimization based formulations facilitate a continuous time solution to the super--resolution problem. Here we treat super--resolution \com{from low--pass measurements} in Phase Space. The Phase Space transformation parametrically generalizes a number of well known unitary mappings such as the Fractional Fourier, Fresnel, Laplace and Fourier transforms. Consequently, our work provides a general super--resolution strategy which is backward compatible with the usual Fourier domain result. We consider low--pass measurements of Dirac distributions in Phase Space and show that the super--resolution problem can be cast as Total Variation minimization. Remarkably, even though are setting is quite general, the bounds on the minimum separation distance of Dirac distributions is comparable to existing methods.}\par\vspace{3mm}}
 \tableofcontents
 \begin{spacing}{1.5}
\section{Introduction}
\label{sec:intro}

Ernst Abbe's foundational work \cite{Abbe1973} in 1873 reported an observation regarding lack of resolvability of optical features beyond the diffraction limit. This problem is central to several areas of science and engineering such as optics\cite{Zheludev2008, Wang2012}, imaging \cite{Hao2013}, geophysics \cite{Levy1981}, depth sensing \cite{Bhandari2014sparse} and astronomy \cite{Puschmann2005}. 

In its abstract form, the problem can be stated as follows: How can we recover a $K$--sparse signal (\com{spike train/Dirac impulses}) with unknown locations and amplitudes from the knowledge of its low--pass measurements in \com{the} Fourier domain? The problem is challenging because it asks for recovery of a non--bandlimited signal from its projection onto \com{the} subspace of bandlimited functions. 
%
%
Two predominant approaches exist in the literature \com{to super--resolution}: optimization and parameter estimation based solutions. 

\vspace{6pt}
\bul \textbf{Optimization Based Super--Resolution:} Early roots of $\ell_1$ norm based optimization can be traced back to \cite{Claerbout73}. 
This \com{approach} has been revitalized due to the advent of \textit{compressed sensing} \cite{EldarG2009}. A very recent idea in this context is that of \textit{continuous sparse modeling} where the signal attributes are estimated on a continuously defined grid \cite{Tang2012,Candes2014} (instead of its discrete counterpart, as is the case for the usual compressive sensing setup).

Bredies and Pikkarainen \cite{Bredies2013} present an optimization method in \com{the} space of signed measures. In their setting,  they consider discrete measurements while the solution space is infinite dimensional. In parallel, de Castro and Gamboa \cite{DeCastro2012} and Cand{\`e}s and Fernandez-Granda \cite{Candes2014} consider a Total Variation (TV) formulation for super--resolution. 
\input{BoundsTable}
%
%
%
%

\vspace{6pt}
\bul \textbf{Parameter Estimation Based Super--Resolution:} The super--resolution problem has been studied in the signal processing context with the goal of resolving overlapping echoes/time--delay estimation \cite{Gedalyahu2010}, multi--path characterization \cite{Bhandari2014sparse} and deconvolution. Li and Speed \cite{Li2000} considered super--resolution in form of parametric deconvolution of spike trains. Vetterli, Blu and co--workers \cite{Blu2008, Vetterli2002} developed the idea of super--resolution as a sparse sampling problem. Eldar and co--workers developed super--resolution methods in the context of sub--Nyquist sampling and \com{the} Xampling framework \com{\cite{Michaeli2012}}. Finally, since the super--resolution problem is closely linked with the spectral estimation problem \cite{Li2000}, MUSIC, ESPIRIT and matrix pencil \cite{Stoica2005, Demanet2013} based methods have also been used. 

\com{
In contrast to parameter estimation techniques where the recovery condition is based on the number of spikes, non--parametric methods provide a bound in the form of a minimum separation condition. More precisely, let $K$ be the number of spikes to be super--resolved, $f_c $ be the cut--off frequency in Fourier domain and let $\Delta$ denote the minimum spacing between any two spikes. While parameter estimation methods require $f_c \geq 2K + 1$ assuming that $K$ is known a priori, non--parametric methods super--resolve the spikes provided that $f_c > f_\textsf{SR}\left( \Delta \right)$ where $f_\textsf{SR}$ is some function. The works of Donoho \cite{Donoho1992}, Kahane \cite{Kahane2011}, Cand{\`e}s and Fernandez-Granda \cite{Candes2014} and Moitra \cite{Moitra2014} provide a theoretical guarantee for the super--resolution problem in terms of $f_\textsf{SR}$. We summarize the recovery guarantees in terms of $f_\textsf{SR}$ in Table~\ref{tab:rg}.}

%
The Fourier transform is well suited for examining signals which are linear combinations of sinusoids. 
In many practical applications such as radar, sonar, holography, wave--physics and quantum optics, the basic building blocks of the signals are not sinusoidal. Often polynomial phase, Fourier--like transformations of form ${e^{\jmath \phi \left( t \right)}}$ are well suited for analysis of such signals. For this purpose, tools such the Fresnel transform \cite{Liebling2003}, Fractional Fourier transform \cite{Almeida1994} and the Chirp transform \cite{Chassande-Mottin1999, Baraniuk1993} have been developed in the literature. 

\com{Our goal here is to extend existing super--resolution results} to integral transformations other than the Fourier transform. 
To this end, we cast the super--resolution problem in phase domain which parametrically generalizes some of the well--known unitary transformations. Some examples are listed in Table~\ref{tab:1}. We show that exact recovery \com{of spike trains from low--pass measurements in phase space} is possible by minimizing the \com{signal's} TV--norm. This is accomplished using a convex program. Remarkably, even with the general construct of the problem, our theoretical guarantee for exact recovery is the same as in \cite{Candes2014}.  
%
%

\com{Throughout the paper}, $\mathbb{R}, \mathbb{C}$ and $\mathbb{Z}$ denote sets of real, complex and integers. We use $z^*$ to denote the complex conjugate of $z\in\mathbb{C}$, $\Re z = \frac{1}{2}\left( {z + {z^*}} \right)$ is the real part of $z$ and $\jmath = \sqrt{-1}$. Discrete sequences are represented by $s\left[ m \right],m \in \mathbb{Z}$ while their continuous counterparts are \com{represented by} $s\left( t \right),t \in \mathbb{R}$. The $L_2$ inner product between functions $f$ and $g$ is denoted by $\left\langle {f,g} \right\rangle  = \int { {f{g^ * }}dt}$. Function composition is denoted by $\left( {f \circ g} \right)\left( x \right) = f\left( {g\left( x \right)} \right)$. \com{Convolution between functions $f$ and $g$ is defined as $\left( {f * g} \right)\left( t \right) = \int {f\left( z \right)g\left( {t - z} \right)dz}$}. We use $\mathbf{s}$ to represent the row--vector of a discrete sequence $s\left[ m \right],m \in [0,\ldots,M-1]$. \com{We use bold capitals to represent matrices, for example $\mathbf{D}$ and $\mathbf{D}^{\dagger}$ is the pseudo--inverse of $\mathbf{D}$. Calligraphic letters such as $\cal{L}$ are used to denote operators. Sets are denoted by capitalized roman font, ${\rm{S}}$. The estimate of function/sequence $\mu$ is represented by $\widetilde \mu$.}

\section{Phase Space Representation of Signals}
The term \textit{Phase Space} naturally arises in areas of optics \cite{Testorf2009} and mathematical physics \cite{Moshinsky1971}. The Fractional Fourier transform (FrFT) \cite{Almeida1994, Bhandari2012,SamilYetik2003,Martone2001} and the Linear Canonical Transform (LCT) \cite{Pei2002,Tao2008,Testorf2009} are instantiations of phase space transformations. These transformations have found a number of applications in signal processing \cite{Almeida1994} and communication problems such as shift--invariant sampling/approximation theory \cite{Bhandari2012,Tao2008}, operator theory, multi--carrier communications \cite{Martone2001}, array processing \cite{SamilYetik2003} and optical signal processing \cite{Testorf2009}. In this paper, we will work with the \textit{Linear Canonical Transform} (LCT) \cite{Moshinsky1971}. 
\begin{definition}[LCT]
Let $\Lambda = \bigl[\begin{smallmatrix}
a&b\\ c&d
\end{smallmatrix} \bigr]$, with $ad-bc = 1$.
The LCT of a function $f\left( t \right), t\in\mathbb{R}$ is a parametric, unitary, integral mapping, ${\mathcal{L}_{\Lambda}} : f \rightarrow \widehat f$ with respect to the time--frequency kernel $k_{\Lambda}$,
\begin{equation}
\label{LCT}
\widehat f\left( \omega  \right) = \underbrace{{\mathcal{L}_{\Lambda}}\left[ f \right]\left( \omega  \right)}_{\textsf{LCT}} \DE 
\begin{cases}
\left\langle {f,k_{\Lambda}\left( { \cdot ,\omega } \right)} \right\rangle & b \neq 0,  \\
\sqrt d {e^{ - \jmath \frac{1}{2}cd{\omega^2}}}f\left( {d\omega} \right) & b =0,
\end{cases}
\end{equation}
where the transformation kernel is defined by
\begin{equation}
\label{kernel}
k_{\Lambda}\left( {t,\omega } \right) = \frac{1}{{\sqrt { - \jmath 2\pi b} }}\exp \left( { - \jmath \frac{1}{{2b}}\left( {a{t^2} + d{\omega ^2} - 2\omega t} \right)} \right).
\end{equation}
\end{definition}
Now since $b=0$ amounts to dilating the function $\sqrt d {e^{ - \jmath \frac{1}{2}cd{\omega^2}}}f\left( {d\omega} \right)$ (see (\ref{LCT})), we will develop our results for the case when $b\neq 0$.

The LCT satisfies a useful operator composition property: $\Ls{1} \circ \Ls{2} = \Ls{3}$ with $\Lambda_3 = \Lambda_2\Lambda_1$. This can be used to show that, 
\begin{equation}
\label{iLCT}
f\left( t \right) = \underbrace{\iL[~ {\widehat f}~]\left( t \right)}_{\textsf{Inverse--LCT}} \DE 
\begin{cases}
\left\langle {\widehat f,k_{\Lambda^{-1}}\left( {t, \cdot } \right)} \right\rangle & b\neq 0\\ 
\sqrt a {e^{ - \jmath \frac{1}{2}ca{t^2}}}f\left( {at} \right) & b=0.
\end{cases}
\end{equation}

The LCT parametrizes a number of well--known, unitary transformations, some of which are listed in Table~\ref{tab:1}. For $\Lambda = \bigl[\begin{smallmatrix} 0 & 1\\ -1 & 0  \end{smallmatrix} \bigr] = \Lambda_{\sf{FT}}$, the LCT amounts to the Fourier transform (upto a constant). (FT). Similarly, with the $2\times 2$ rotation matrix $\Lambda_{\theta}$ (see Table~\ref{tab:1}), we obtain the Fractional Fourier transform (FrFT). 
\input{Table1LCT}
Matrix factorization shows that the LCT is related to the FT and the FrFT. The Fourier matrix factorization is simple: 
\[{\footnotesize\left[ {\begin{array}{*{20}{c}}
  a&b \\ 
  c&d 
\end{array}} \right] = \left[ {\begin{array}{*{20}{c}}
  b&0 \\ 
  d&{{b^{ - 1}}} 
\end{array}} \right]\Lambda_{\sf{FT}} \left[ {\begin{array}{*{20}{c}}
  1&0 \\ 
  {a/b}&1 
\end{array}} \right] \Leftrightarrow \Lambda  = \mathbf{M}_1\Lambda_{\sf{FT}} \mathbf{M}_2,}\]
and leads to the implementation ${\mathcal{L}_{\Lambda}} = {\mathcal{L}}_{{\mathbf{M}}_1} \circ \Ls{\sf{FT}} \circ  {\mathcal{L}}_{{\mathbf{M}}_2}$. 
The Fractional Fourier matrix factorization is more involved. Relying on the \textit{Iwasawa Decomposition} \cite{Iwasawa1949}, 
\[{\footnotesize \left[ {\begin{array}{*{20}{c}}
  a&b \\ 
  c&d 
\end{array}} \right] = \Lambda_{\theta} \left[ {\begin{array}{*{20}{c}}
  \Gamma &0 \\ 
  0&{{\Gamma ^{ - 1}}} 
\end{array}} \right]\left[ {\begin{array}{*{20}{c}}
  1&u \\ 
  0&1 
\end{array}} \right] \Leftrightarrow \Lambda  = \Lambda_{\theta} \mathbf{DU},}\]
where $\mathbf{D}$ is a diagonal matrix with $\Gamma  = \sqrt{a^2 + c^2}$ and $\mathbf{U}$ is an upper--triangular matrix with $u = \left( ab+cd \right) / \Gamma^2$.

A useful operation that is linked with phase space is the convolution/filtering operator. For the Fourier domain, we have the convolution--multiplication property: $\left( {f * g} \right) =$ $\iL\left[ {{\mathcal{L}_{\Lambda}} \left[ f \right] {\mathcal{L}_{\Lambda}} \left[ g \right]} \right],$ with $\Lambda = \Lambda_\textsf{FT}$. Unfortunately, this property is not preserved in phase space. To circumvent this problem, we use a version of the FrFT convolution operator \cite{Bhandari2012}.
\begin{definition}[LCT Convolution/Filtering]
Let $*_{\Lambda}$ denote the convolution/filtering operation in LCT domain and $*$ be the usual Fourier domain convolution operator. Convolution of functions $f$ and $g$ in the LCT domain is defined by  
\begin{equation}
\label{LCTconv}
\left( {f{*_{\Lambda}}g} \right)\left( t \right) = \frac{{{e^{ - \jmath \frac{{a{t^2}}}{{2b}}}}}}{{\sqrt {\jmath 2\pi b} }}
\underbrace{\left( {f\left( t \right){e^{ + \jmath \frac{{a{t^2}}}{{2b}}}}*g\left( t \right){e^{ + \jmath \frac{{a{t^2}}}{{2b}}}}} \right)}_{\textsf{Convolution of Modulated Functions}}.
\end{equation}
\end{definition}
By following the steps in \cite{Bhandari2012}, it is easy to verify that the operation in (\ref{LCTconv}) admits a convolution--multiplication property:  
\begin{equation}
\label{duality}
{\mathcal{L}_{\Lambda}}\left[ {f{*_{\Lambda} }g} \right] \left( \omega \right)= {e^{ - \jmath \frac{{d{\omega ^2}}}{{2b}}}}\widehat f\left( \omega  \right)\widehat g\left( \omega  \right).
\end{equation}

\section{Sparse Signals in Phase Space}
%
Consider a $K$--sparse object/spike train modeled by, 
\begin{equation}
\label{sod}
s\left( t \right) = \sum\nolimits_{k = 0}^{K - 1} {{c_k}\delta \left( {t - {t_k}} \right),\ t \in \mathbb{R}} 
\end{equation}
where $\delta$ denotes the Dirac mass with weights $\{c_k \}_k \in \mathbb{C}$ that is activated on locations $\{t_k\} \in \left[ 0 , \tau \right), k = 0,\ldots,K-1$. Since we are dealing with a finite length signal $s$ that lives on the interval $\left[ 0 , \tau \right)$, we investigate its representation in phase space using the Fractional Fourier series \cite{Pei1999} analog of the LCT. 

\begin{definition}[Linear Canonical Series] Let $f$ be a compactly supported function such that \com{$f\neq 0, \forall t \in \left[0, \tau \right)$ and zero elsewhere} \com{and let ${\kappa _{b,\tau }} = \sqrt {2\pi b/\tau }$, $b\neq0$}. The Linear Canonical Series (LCS) \com{expansion} of the function $f$ is given by 
\begin{equation}
\label{LCS}
f\left( t \right) = {\kappa _{b,\tau }} \sum\limits_{n =  - \infty }^{n =  + \infty } { \widehat{f}\left[n \right]{k_{\Lambda}} \left( {t,n{\omega _0}b} \right)}, \quad \omega_0 = \frac{2\pi}{\tau}. 
\end{equation}
The LCS coefficients $\widehat{f}$ are evaluated at $\omega = n \omega_0 b, n\in \mathbb{Z}$, 
\begin{equation}
\label{LCC}
\widehat f\left[ n \right] =  {\kappa _{b,\tau }} {\mathcal{L}_{\Lambda}}\left[ f \right]\left( {n{\omega _0}b} \right) \equiv {\kappa _{b,\tau }} {\left\langle {f,{k_{\Lambda} }\left( {,n{\omega _0}b} \right)} \right\rangle }.
\end{equation}
\end{definition}
Note that by appropriately parameterizing $\Lambda$, one can easily design the basis functions for any of the transformations listed in Table~\ref{tab:1}. \com{For example, with ${\Lambda} = \Lambda_{\textsf{FT}}$ in (\ref{LCC}), we obtain the Fourier Series expansion of $f$.} 

We now compute the series coefficients for the sparse signal $s\left(t \right)$ in (\ref{sod}). We use (\ref{LCC}), to compute the LCS coefficients
\begin{align}
  \widehat s\left[ n \right] = {\kappa _{b,\tau }} {\mathcal{L}_{\Lambda}}\left[ f \right]\left( {n{\omega _0}b} \right) 
&   \EQc{(\ref{LCC})} {\kappa _{b,\tau }} \int_\tau  {s\left( t \right) k_{\Lambda} ^*\left( {t,n{\omega _0}b} \right) dt}  \notag \hfill \\
&   = {\kappa _{b,\tau }} \sum\nolimits_{k = 0}^{K - 1} {{c_k}k_{\Lambda} ^*\left( {{t_k},n{\omega _0}b} \right)}.
\label{sparseLCS}
\end{align}
Plugging the coefficients into the LCS representation (\ref{LCS}), we \com{obtain} the phase space representation of $s\left( {t} \right)$ 
\begin{align}
  s\left( t \right) & \EQc{(\ref{LCS})} {\kappa _{b,\tau }} \sum\limits_{m =  - \infty }^{m =  + \infty } {\widehat{s}\left[ m \right] {k_{\Lambda} }\left( {t,m{\omega _0}b} \right)}, \quad \omega_0 = \frac{2\pi}{\tau} \notag  \hfill \\
  & = \frac{{{e^{ - \jmath \frac{{a{t^2}}}{{2b}}}}}}{{\tau}}\sum\limits_{m =  - \infty }^{m =  + \infty } {\underbrace {\left( {\sum\limits_{k = 0}^{K - 1} {{c_k}{e^{\jmath \frac{a}{{2b}}t_k^2}}{e^{ - \jmath m{\omega _0}{t_k}}}} } \right)}_{\widehat y\left[ m \right]}{e^{\jmath m{\omega _0}t}}} \label{ym} \hfill \\
  & = \frac{{{e^{ - \jmath \frac{{a{t^2}}}{{2b}}}}}}{{\tau}}\sum\limits_{m =  - \infty }^{m =  + \infty } {\widehat y\left[ m \right]{e^{\jmath m{\omega _0}t}}}. 
\label{SOE}
\end{align}
%
From \cite{Candes2014}, we see that $\widehat{y} \left[m\right]$ are precisely the Fourier series coefficients of $\tau s\left( t \right){e^{\jmath \frac{{a{t^2}}}{{2b}}}}$. Thus, even though we are dealing with the phase space, the linear frequency modulated, sparse signal $s$ is completely characterized by its Fourier series coefficients $\widehat y$. 

\section{Super--Resolution in Phase Space}
\subsection{Problem Formulation}
Let $\operatorname{sinc} \left( t \right) = \frac{{\sin \left( {\pi t} \right)}}{{\pi t}}$. Consider the frequency modulated function, 
\begin{equation}
{\phi _{{\textsf{LP}}}}\left( t \right) = \left( {\Omega /b} \right){e^{ - \jmath \frac{{a{t^2}}}{{2b}}}}\operatorname{sinc} \left( {\left( {\Omega /b} \right)t} \right).
\label{sinc}
\end{equation}
Note that this function is $\left(\Omega\pi\right)$--bandlimited because in phase space the function $\phi _{\textsf{LP}}\left( t \right)$ is compactly supported, or, 
$${\widehat \phi _{{\textsf{LP}}}}\left( \omega  \right) = {\mathcal{L}_{\Lambda}}[{\phi_{\textsf{LP}}}]\left(\omega\right) =  \frac{{{e^{ + \jmath \frac{{d{\omega ^2}}}{{2b}}}}}}{{\sqrt {\jmath 2\pi b} }}\Pi \left( {\frac{{\omega}}{{2\pi \Omega }}} \right)$$ 
where $\Pi \left( \omega  \right) = 1,\left| \omega  \right| \leqslant 1/2$ and zero, elsewhere.

With $s$ defined in (\ref{SOE}), its low--pass version is, 
\begin{equation}
h\left( t \right) = \underbrace {\left( {s{*_{\Lambda} }{\phi _{{\textsf{LP}}}}} \right)\left( t \right)}_{{\textsf{Low--Pass Filtering}}} \EQc{(\ref{LCTconv})} \frac{{{e^{ - \jmath \frac{{a{t^2}}}{{2b}}}}}}{{\sqrt {\j 2\pi b} }}\frac{1}{{ \tau }}\sum\limits_{\left| m \right| \leqslant \left\lfloor {\Omega \tau/2b} \right\rfloor } {\widehat y\left[ m \right]{e^{\jmath {\omega _0}mt}}},
\label{LPMT}
\end{equation}
where $\left\lfloor  \cdot  \right\rfloor$ is the floor operation. 

Suppose we observe $N$ discrete, low--pass measurements sampled with sampling rate $T = b/\Omega$, 
\begin{equation}
h\left[ n \right] = {\left. {h\left( t \right)} \right|_{t = nT}},\ \ T = b/\Omega, \ \ n = 0,\ldots, N-1.
\label{LPM}
\end{equation}
By modulating $h$, we obtain measurements of the form, 
\begin{equation}
y\left[ n \right] = \underbrace {\sqrt {\j 2\pi b \tau} {e^{ + \jmath \frac{{a{\left(nT\right)^2}}}{{2b}}}}h\left[ n  \right]}_{{\textsf{Modulated Measurements}}} = \sum\limits_{\left| m \right| \leqslant {f_c}} {\widehat y\left[ m \right]{e^{\jmath {\omega _0}m \left( nT \right)}}} 
\label{yhat}
\end{equation}
where ${f_c} = \left\lfloor {\Omega \tau/2b} \right\rfloor$. In vector--matrix notation, we have low--pass measurements, $ {\mathbf{y}} = \idft\widehat {\mathbf{y}}$, where $\idft \in \mathbb{C}^{N\times \left(2f_c-1\right)}$ is the usual inverse--DFT matrix with elements ${\left[ \idft \right]_{n,m}} = {e^{\jmath m{\omega _0}\left(nT\right)}}$ and we assume that $N \geqslant 2{f_c} - 1$ so that $\idft$ is a full--rank marix.

Having obtained $\widehat{\mathbf{y}}$, the question then is: how many samples of $\widehat{\mathbf{y}}$ are sufficient for complete characterization of $\widehat{y}\left[ m \right] \EQc{(\ref{ym}) } \sum\nolimits_{k = 0}^{K - 1} {\left( {{c_k}{e^{ + \jmath \frac{a}{{2b}}t_k^2}}} \right){e^{ - \jmath m{\omega _0}{t_k}}}}$? Indeed, from spectral estimation theory \cite{Stoica2005}, we know that we need at least $2K+1$ values of $\widehat{y}$ to solve for $\{c_k,t_k\}_k$. Consequently, whenever
\begin{equation}
{f_c} = \left\lfloor {\tfrac{{\Omega \tau }}{{2b}}} \right\rfloor  \geqslant K \Leftrightarrow \left\lfloor {\tfrac{\tau }{{2T}}} \right\rfloor  \geqslant K,
\label{samplingrate}
\end{equation}
the system of equations is complete in $\widehat{y}$ meaning that we have at least $2K+1$ values of $\widehat{y}$ and we can solve for $\{c_k,t_k\}$. Thus (\ref{samplingrate}) provides a bound on the minimum sampling density in phase space. 

\subsection{Super--Resolution Via Convex Programming}
Given $N$ measurements $\mathbf{y}$, we obtain $\widehat{\mathbf{y}}$ using $\widehat{\mathbf{y}} = \idft^{\dagger}{\mathbf{y}}$ (\ref{yhat}). From the phase space development of the problem, we know that
\begin{align}
  \widehat y\left[ m \right] & \EQc{(\ref{ym})} \sum\nolimits_{k = 0}^{K - 1} {\left( {{c_k}{e^{ + \jmath \frac{a}{{2b}}t_k^2}}} \right){e^{ - \jmath m{\omega _0}{t_k}}}}  \label{constraints}  \\
 &  =  \int_0^\tau  { \mu \left( t \right) {e^{ - \jmath m{\omega _0}t}} dt }, \ \ \left| m \right| \leqslant  {f_c} = \left\lfloor {\Omega \tau/2b} \right\rfloor  \notag\hfill 
 \end{align}
%
%
where
\begin{equation}
\label{mu}
\mu \left( t \right) = \sum\nolimits_{k = 0}^{K - 1} {\underbrace {{c_k}{e^{ + \jmath \frac{a}{{2b}}t_k^2}}}_{{\rho _k}}\delta \left( {t - {t_k}} \right)}
\end{equation}
and ${\rho _k} \DE {c_k}{e^{ + \jmath \frac{a}{{2b}}t_k^2}}$ are the new weights for $s\left( t\right)$. 

We are now left to solve the standard super--resolution problem \cite{Candes2014} where one has access to the low--pass measurements $\widehat y\left[ m \right]$ and the signal to be super--resolved is prescribed in (\ref{mu}). Hence, the problem of recovering $\mu$ from $\widehat{\mathbf{y}}$ can now be solved by using, 
%
%
\begin{empheq}[box={\Yellowbox[\color{red} \scriptsize \textsf{Super--Resolution in Phase Space (Primal Problem)}]}]{equation}
\mathop {\min }\limits_{\widetilde \mu } {\left\| {\widetilde \mu } \right\|_{{\textsf{TV}}}} \ \ {\text{subject to}} \ \  {\left\{ {\widehat y\left[ m \right] \EQc{(\ref{constraints})} \int_0^\tau  {\widetilde \mu \left( t \right) {e^{ - \jmath m{\omega _0}t}} dt} } \right\}_{\left| m \right| \leqslant {f_c}}
\label{CP}}
\end{empheq}
where, for the model assumed in (\ref{mu}), the \textsf{TV}--norm amounts to, ${\left\| { \mu } \right\|_{{\text{TV}}}} = \sum\nolimits_k {\left| {{\rho _k}} \right|}  \EQc{(\ref{mu})} \sum\nolimits_k {\left| {{c_k}} \right|}.$ We note that in principle ${\left\| {\widetilde \mu } \right\|_{{\textsf{TV}}}} = {\left\| s \right\|_{{\textsf{TV}}}}$, however, due to the nature of phase space measurements, the real/complex weights $\{c_k\}_k$ need to be demodulated using the linear frequency modulation term ${e^{ + \jmath \frac{a}{{2b}}t_k^2}}$ which depends on $\Lambda$.

The problem in (\ref{CP}) seeks to recover the infinite--dimensional variable $\widehat\mu$ from finitely many constraints set up in (\ref{constraints}). This continuous optimization problem has a tractable dual problem. As was shown in \cite{Candes2014}, a semidefinite program (SDP) can be used to recover $\widetilde\mu$ by computing $\{t_k \}_k$ first and then recovering $\{c_k\}_k$ using a least squares fit. 
%
The SDP equivalent \cite{Candes2014}  of the convex dual of (\ref{CP}) is,
\begin{empheq}[box={\Yellowbox[\color{red} \scriptsize \textsf{Semidefinite Program (Dual Problem)}]}]{align*}
 & \mathop {\max }\limits_{{\mathbf{u}},{\mathbf{M}}} \ \ \ \Re \left\langle {\widehat {\mathbf{y}},{\mathbf{u}}} \right\rangle \qquad  {\text{subject to,}} \hfill \\
&  \left[ {\begin{array}{*{20}{c}}
  {\mathbf{M}}&{\mathbf{u}} \\ 
  {{{\mathbf{u}}^*}}&1 
\end{array}} \right] \pmb{\succ} 0,\quad \sum\nolimits_{j \in {\rm{S}_2},m \in {\rm{S}_1}} {{{\left[ {\mathbf{M}} \right]}_{m,m + j}}}  = {\delta _j} \notag
\end{empheq}
where ${\rm{S}_1} = \left[ {1,2{f_c} + 1 - j} \right]$, ${\rm{S}_2} = \left[ {0,2{f_c}} \right]$,  $\mathbf{M} \in \mathbb{C}^{\left(2f_c+1\right) \times \left(2f_c+1\right)}$ is some Hermitian matrix and $\mathbf{u} \in \mathbb{C}^{2f_c + 1}$ is a complex vector. 

The SDP input $\widehat{\mathbf{y}}$ results in a vector $\mathbf{u}$. In order to recover the locations $\{t_k\}_k$, we construct the polynomial of degree $N_0=4f_c$, 
\begin{equation}
{p_{N_0}}\left( z \right) = 1 - \sum\nolimits_{ | k | \leq 2f_c } {{u_k}{z^k}}, \quad z\in \mathbb{C}.
\label{poly}
\end{equation}
The roots of $p_{4f_c}\left( z \right), z = {e^{\jmath {\omega _0}t}} $ lead to the locations $\{t_k\}_k$. Knowing $\widehat{\mathbf{y}}$ together with the estimates, $\{\widetilde t_k\}_k$, we use the constraints in (\ref{constraints}) to set up a system of equations which leads to amplitude estimates $\widetilde{c_k} = {\widetilde\rho _k}{e^{ - \jmath \frac{a}{{2b}} \widetilde{t}_k^2}}.$ (see (\ref{mu})). 
Finally, we recover our super--resolved signal, $\widetilde s\left( t \right) = \sum\nolimits_{k = 0}^{K - 1} {{{\widetilde c}_k}\delta \left( {t - {{\widetilde t}_k}} \right)}.$
Stepwise procedure for super--resolution in phase space is outlined in Algorithm 1.

In view of \cite{Candes2014}, let us invoke the definition of minimum distance $\Delta  = {\inf _{{{\left\{ {{t_k}} \right\}}_k}:{t_k} \ne {t_l}}}\left| {{t_k} - {t_l}} \right|$. With ${f_c} = \left\lfloor {\Omega \tau /2b} \right\rfloor$ , the exact recovery requirement for phase space is as follows:
\begin{theorem}[Exact Recovery in Phase Space]
Let the support set of $s\left( t \right) $ in (\ref{sod}) be ${\rm{S}} = \left\{ {{{\widetilde t}_k}} \right\}_k$. If the minimum distance obeys the bound $\Delta \left( {\rm{S}} \right){f_c} \geqslant 2$, then $s\left( t \right) $ is a unique solution to (\ref{CP}).
\end{theorem}
The proof of this theorem is a straight--forward consequence of \cite{Candes2014}. Moreover, due to inherent Fourier structure of the phase space problem, our work may benefit from the ideas discussed in \cite{Donoho1992, DeCastro2012,Moitra2014,Demanet2013}.

\subsection{Remarks and Discussion}
\bul \textbf{Backward Compatibility} With the choice of parameter matrix $\Lambda = \Lambda_{\textsf{FT}}$ (cf. Table~\ref{tab:1}), our result coincides with the usual, Fourier domain case of super--resolution \cite{Candes2014, DeCastro2012}. Furthermore, $\Lambda = \Lambda_{\theta}$ relates to the case of Fractional Fourier domain for which our result generalizes a previous known result \cite{Bhandari2010}.

\bul \textbf{Exact Recovery Condition} Even though our super--resolution naturally extends to a number of well known unitary transformations, the exact recovery condition remains unchanged. Hence re-formulating the super--resolution problem in context of phase space comes at no extra cost in the sense of recovery requirement. 
%
\input{AlgorithmPSSR}

\subsection{An Application of Super--Resolution in Phase Space}
Bandlimted signals are compactly supported in the Fourier domain. When a bandlimited signal is corrupted by additive impulsive noise or AIN, the holes/zeros in the spectrum are filled by the spectral components that characterize the impulsive noise which is essentially non--bandlimited. Wolf \cite{Wolf1983} used the idea of curve--fitting the out--of--band components for identification of impulsive noise components. Here, we formulate the problem of denoising linear frequency modulated (LFM) signals that are corrupted by AIN. Since LFM signals are the basis functions of phase space transformations, it is clear that such signals are bandlimited in the LCT domain. 

Consider a bandlimited LFM signal $${r_{{\textsf{BL}}}}\left( t \right) = {\kappa _{b,\tau }}\sum\nolimits_{|m| \leq M} {{{\widehat r}_{{\textsf{BL}}}}\left[ m \right]{k_{\Lambda} }\left( {t,m{\omega _0}b} \right)},$$ with ${{{\widehat r}_{{\textsf{BL}}}}\left[ m \right]}= 0, |m| > M$ and let $r\left( t \right) = {r_{{\textsf{BL}}}}\left( t \right) + s\left( t \right)$ be the signal corrupted by AIN. Clearly, $r\left( t \right)$ is non--bandlimited in phase space due to $s\left( t \right)$. Suppose we observe low--pass filtered samples of $r\left( t \right)$, that is, 
\[\underbrace {\left( {r{*_{\Lambda}}{\phi _{{\textsf{LP}}}}} \right)\left( {nT} \right)}_{h\left[ n \right]} = {e^{ - \jmath \frac{{a{{\left( {nT} \right)}^2}}}{{2b}}}}\sum\limits_{\left| m \right| \leqslant {f_c}} {\underbrace {\left( {{c_1}{{\widehat y}_1}\left[ m \right] + {c_2}{{\widehat y}_2}\left[ m \right]} \right)}_{{{\widehat y}_r}\left[ m \right]}{e^{\jmath {\omega _0}mnT}}} \]
where $c_1,c_2$ are known constants, ${\widehat y_1}\left[ m \right] = {\widehat r_{{\textsf{BL}}}}\left[ m \right]{e^{ - \jmath \frac{{d{{\left( {m{\omega _0}b} \right)}^2}}}{{2b}}}}$ and ${\widehat y_2}\left[ m \right] = \widehat y\left[ m \right]$ (as in (\ref{constraints})) where ${f_c} = \left\lfloor {\Omega \tau/2b} \right\rfloor \equiv \left\lfloor { \tau/2 T} \right\rfloor $. Again, let us define $y\left[ n \right] = h\left[ n \right]{e^{\jmath \frac{{a{{\left( {nT} \right)}^2}}}{{2b}}}}, n = 0,\ldots,N-1, N\geq 2f_c +1$. Provided that $f_c \geq M+2K+1$, we have, 
\[ 
\widehat{y}_r \left[ m \right ] = 
\begin{cases}
{c_1}{\widehat y_1}\left[ m \right] + {c_2}{\widehat y_2}\left[ m \right] & |m| \leq M \\
{c_2}{\widehat y_2}\left[ m \right] & \left| m \right| > M 
\end{cases},
\]
which leads to complete characterization of $s\left(t\right)$ since the $2K+1$ values of $\widehat{y_2}\left[ m \right]$ can be used with (\ref{CP}) to solve for $s\left( t \right)$. With $\widehat{\mathbf{y}} = \widehat{\mathbf{y}}_r $, $m>M$ we can use Algorithm 1 for exact denoising of $r\left( t \right)$.

\section{Conclusion}
We develop a method for super--resolution in phase space. The phase space transformation generalizes a number of well known transforms (see Table~\ref{tab:1}). More precisely, we are concerned with recovery of spike trains from their low--pass samples. For this purpose, we filter the spike train with a kernel which is bandlimited in phase space. We show that even though we are dealing with a general class of parametric transformations, the low--pass samples are completely characterized by chirp--modulated Fourier series. Having made this link, we show that the recovery of spikes from their low--rate measurements can be cast as a total--variation minimization---a problem that can be tackled by convex programming. In closing, our work extends the recent results of \cite{Candes2014} without altering the exact recovery condition. That said, the cut--off frequency is a function of the transform being used for investigation. Our work warrants future research, specially for the case of additive noise. 
\end{spacing}
\begin{spacing}{1.5}
\bibliographystyle{IEEEbib}
\bibliography{IEEEabrv,SuperRes,ChirpBiblio}
\end{spacing}

\end{document}

%% file: ABPreICASSP2k15.tex
\usepackage{array,graphicx}
\usepackage{amsfonts,cite}
\usepackage[cmex10]{amsmath}
\usepackage{mathtools}
\usepackage[amsmath]{empheq}
\usepackage{cases,balance,cite, lineno}
\usepackage{booktabs,mathrsfs,bbm,amssymb,soul}
\usepackage[table]{xcolor}

\usepackage{caption}
\usepackage{setspace}

\definecolor{myblue}{rgb}{0,0,1}

\newcommand{\bul}{\noindent {\color{red}\ensuremath{\blacksquare} }}

\let\OldSymbol\Delta
\renewcommand{\Delta}{\pmb{\OldSymbol{}}}

\let\OldLambda\Lambda
\renewcommand\Lambda{\boldsymbol{\OldLambda{}}}

\def\iL{\mathcal{L}_{{\Lambda}^{-1}}}
\def\j{\jmath}
\def\DE{\stackrel{\mathrm{def}}{=}}

\def\idft{\mathbf{V}_{\textsf{IDFT}}}

\newtheorem{theorem}{Theorem}
\newtheorem{definition}{Definition}

\newcommand{\Ls}[1]{\mathcal{L}_{{\Lambda}_{#1}}}
\newcommand{\EQc}[1]{\stackrel{{#1}}{=}}
\newcommand{\com}[1]{{\color{black}{#1}}}

%% file: BoxedEquations.tex
\usepackage{empheq}

\definecolor{shadecolor}{rgb}{0.96,0.96,0.96}
\definecolor{light-blue}{rgb}{1,1,1}
\newsavebox{\mysaveboxM}
\newsavebox{\mysaveboxT}

\newcommand*\Yellowbox[2][Formel]{%
  \sbox{\mysaveboxM}{#2}%
  \sbox{\mysaveboxT}{\fcolorbox{black}{light-blue}{#1}}%
  \sbox{\mysaveboxM}{%
    \parbox[b][\ht\mysaveboxM+0.4\ht\mysaveboxT+0.8\dp\mysaveboxT][b]{%
      \wd\mysaveboxM}{#2}%
  }
  \sbox{\mysaveboxM}{%
    \fcolorbox{black}{shadecolor}{%
      \makebox[\displaywidth-2\fboxsep-2\fboxrule]{\usebox{\mysaveboxM}}%
    }%
  }%
  \usebox{\mysaveboxM}%
  \makebox[0pt][r]{%
    \makebox[\wd\mysaveboxM][c]{%
      \raisebox{\ht\mysaveboxM-0.4\ht\mysaveboxT
                +0.5\dp\mysaveboxT-0.5\fboxrule}{\usebox{\mysaveboxT}}%
    }%
  }%
}

%% file: BoundsTable.tex
{\begin{table}[t]
\centering
\begin{tabular*}{0.7\textwidth}{c*{4}{c}}

\multicolumn{4}{c}{ $\Delta$: Minimum separation $f_c$: Cut--off frequency}  \tabularnewline
\midrule
\hspace{-0pt}Donoho \cite{Donoho1992} & Kahane  \cite{Kahane2011} & Cand{\`e}s--G. \cite{Candes2014} & Moitra \cite{Moitra2014}\tabularnewline [10pt]
\hspace{-0pt} $\Delta > \frac{1}{f_c}$ & $\Delta>\frac{5}{{{f_c}}}\sqrt {\log \left( \frac{1}{2f_c}\right)}$ & $\Delta > \frac{2}{f_c}$ & $\Delta> \frac{1}{f_c-1}$ \tabularnewline  
\bottomrule
\tabularnewline
\end{tabular*}
\caption{Exact Recovery Condition for Super--Resolution.}
\label{tab:rg}
\end{table}}

%% file: Table1LCT.tex
%
\begin{table}[!t]
\centering
\caption{ \textrm{Parametric Representation of Unitary Transformations}}
\begin{tabular*}{1\textwidth}{b{10cm} b{5.2cm}}
\midrule
\rowcolor{blue!10}
\textsf{Parameter Matrix} $\left({\Lambda} \right) $ & \textsf{Corresponding Transform} \\
\midrule
$\bigl[ \begin{smallmatrix} \cos\theta&\sin\theta \\
 -\sin\theta&\cos\theta\end{smallmatrix} \bigr] = \Lambda_\theta $				&  \textbf{Fractional Fourier Transform} \\	[3pt] 
$\bigl[ \begin{smallmatrix} 0&1\\-1&0\end{smallmatrix} \bigr] 
= \Lambda_\textsf{FT}$  													& \textbf{Fourier Transform (FT)}  \\		[3pt] 
$\bigl[ \begin{smallmatrix} 0& \j \\ \j&0\end{smallmatrix} \bigr] 
= \Lambda_\textsf{LT}$ 													& \textbf{Laplace Transform (LT)}  \\	[3pt] 
$\bigl[ \begin{smallmatrix} \j  \cos\theta& \j \sin\theta \\
 \j \sin \theta & -\j \cos\theta \end{smallmatrix} \bigr] $  								& \textbf{Fractional Laplace Transform}   \\ 	[3pt] 
$\bigl[ \begin{smallmatrix} 1&b\\0&1\end{smallmatrix} \bigr]$  						& \textbf{Fresnel Transform}  \\	[3pt] 
$\bigl[ \begin{smallmatrix} 1&\jmath b \\  \jmath&1\end{smallmatrix} \bigr]$				& \textbf{Bilateral Laplace Transform}  \\	[3pt] 
$\bigl[ \begin{smallmatrix} 1&-\jmath b \\ 0&1\end{smallmatrix} \bigr]$, $b \ge 0$  		& \textbf{Gauss--Weierstrass Transform}   \\	[3pt] 
$\tfrac{1}{{\sqrt 2 }} \bigl[\begin{smallmatrix} 0 & e^{ - {{\jmath\pi } 
\mathord{\left/{\vphantom {{j\pi } 2}} \right.\kern-\nulldelimiterspace} 2}} 
\\-e^{ - {{\jmath\pi } \mathord{\left/{\vphantom {{j\pi } 2}} \right.\kern-\nulldelimiterspace} 2}} 
&1\end{smallmatrix} \bigr]$  													& \textbf{Bargmann Transform} \\[3pt] 
\bottomrule
\end{tabular*}
\label{tab:1}
\end{table}

%% file: AlgorithmPSSR.tex
\begin{algorithm}[!t]
 \caption{Super--Resolution in Phase Space.}
\footnotesize
 \textbf{Input:} Low-pass samples $h\left[n\right] \EQc{(\ref{LPMT})} \left( s *_{\Lambda} \phi_\textsf{LP} \right) \left( nT \right)$\\ 
 \textbf{Modulate Samples} $h\left[ n \right] \to \sqrt {\j 2\pi b\left| \tau  \right|} {e^{ + \jmath \frac{{a{\left(nT\right)^2}}}{{2b}}}}h\left[ n \right] = y\left[ n \right]$\\ 
 \KwData{$\widehat{\mathbf{y}} = \idft^{\dagger} \mathbf{y}, \widehat{\mathbf{y}} \in \mathbb{C}^{2f_c + 1}$}
 \textbf{Solve SDP :} \ \ \ \  $\mathop {\max }\limits_{{\mathbf{u}},{\mathbf{M}}} \ \Re \left\langle {\widehat {\mathbf{y}},{\mathbf{u}}} \right\rangle \ \  {\text{subject to}} \ \ \left[ {\begin{array}{*{20}{c}}
  {\mathbf{M}}&{\mathbf{u}} \\ 
  {{{\mathbf{u}}^*}}&1 
\end{array}} \right] \succ 0$
 
 \textbf{Construct Polynomial: }  ${p_{{N_0}}}\left( z \right) = 1 - \sum\nolimits_{\left| k \right| \leqslant 2{f_c}} {{u_k}{z^k}}$ 

 \textbf{Support: } ${p_{{N_0}}}\left( {{e^{\jmath {\omega _0}t}}} \right) = 0 \to \left\{ {{{\widetilde t}_k}} \right\}_k$

 \textbf{Weights: } 
  $\mathop {\min }\limits_{{{\widetilde \rho }_k}} {\left| {\widehat y\left[ m \right] - \sum\nolimits_{k = 0}^{K - 1} {{{\widetilde \rho }_k}{e^{ - \jmath {\omega _0}{{\widetilde t}_k}}}} } \right|^2} \to \{\widetilde \rho_k\}_k$
 

\textbf{Output:} $\widetilde{s} \left( t \right) = \sum\nolimits_{k = 0}^{K - 1} {{\widetilde{c}_k}\delta \left( {t - {\widetilde{t}_k}} \right)}$, $\{{\widetilde c_k} = {\widetilde \rho _k}{e^{ - \jmath \frac{a}{{2b}}\widetilde t_k^2}} \}_k$
\end{algorithm}